\documentclass[lettersize,journal]{IEEEtran}

\usepackage{amsmath,amsfonts}
\usepackage{algorithmic}
\usepackage{array}
\usepackage[caption=false,font=normalsize,labelfont=sf,textfont=sf]{subfig}
\usepackage{textcomp}
\usepackage{stfloats}
\usepackage{url}
\usepackage{verbatim}
\usepackage{graphicx}
\usepackage{cite}
\usepackage{graphicx}
\usepackage{epstopdf}
\usepackage{amssymb}
\usepackage[ruled, linesnumbered]{algorithm2e}

\usepackage{multirow} 

\usepackage{amsmath}

\usepackage{xcolor}

\newtheorem{theorem}{Theorem}
\newtheorem{assumption}{Assumption}
\newtheorem{remark}{Remark}
\begin{document}

\title{Resource Efficient Asynchronous Federated Learning for Digital Twin Empowered IoT Network}

\author{Shunfeng~Chu,
        Jun~Li, 
        Jianxin~Wang,
        Yiyang~Ni,
        Kang~Wei, 
        Wen~Chen,
        Shi~Jin
%
\thanks{S. Chu, and J. Wang are with the School of Electronic and Optical Engineering, Nanjing University of Science and Technology, Nanjing 210094, China (e-mail: shunfeng.chu@njust.edu.cn; wangjxin@njust.edu.cn).}
\thanks{J. Li is with the School of Information Science and Engineering, Southeast University, Nanjing 211189, China (e-mail: jleesr80@gmail.com).}
\thanks{Y. Ni is with the Jiangsu Second Normal University, also with Jiangsu Key Laboratory of Wireless Communications, Nanjing University of Posts and Telecommunications, Naniing 210003. China(e-mail: niyy@jssnu.edu.cn).}
\thanks{K. Wei was with the School of Electrical and Optical Engineering, Nanjing University of Science and Technology, Nanjing 210094, China. He is now with the Department of Computing, The Hong Kong Polytechnic University, Hong Kong 999077, China. (e-mail: kangwei@polyu.edu.hk).}
\thanks{W. Chen is with the Department of Electronic Engineering, Shanghai Jiao Tong University, Shanghai 200240, China (e-mail: wenchen@sjtu.edu.cn).}
\thanks{S. Jin is with the National Mobile Communication Research Laboratory, Southeast University, Nanjing 210096, China (e-mail: jinshi@seu.edu.cn).}}



\markboth{Draft}%
{Shell \MakeLowercase{\textit{et al.}}: Bare Demo of IEEEtran.cls for IEEE Journals}

\maketitle

\begin{abstract}
As an emerging technology, digital twin (DT) can provide real-time status and dynamic topology mapping for Internet of Things (IoT) devices. However, DT and its implementation within industrial IoT networks necessitates substantial, distributed data support, which often leads to ``data silos'' and raises privacy concerns. To address these issues, we develop a dynamic resource scheduling algorithm tailored for the asynchronous federated learning (FL)-based lightweight DT empowered IoT network. Specifically, our approach aims to minimize a multi-objective function that encompasses both energy consumption and latency by optimizing IoT device selection and transmit power control, subject to FL model performance constraints. We utilize the Lyapunov method to decouple the formulated problem into a series of one-slot optimization problems and develop a two-stage optimization algorithm to achieve the optimal transmission power control and IoT device scheduling strategies. In the first stage, we derive closed-form solutions for optimal transmit power on the IoT device side. In the second stage, since partial state information is unknown, e.g., the transmitting power and computational frequency of IoT device, the edge server employs a multi-armed bandit (MAB) framework to model the IoT device selection problem and utilizes an efficient online algorithm, namely the client utility-based upper confidence bound (CU-UCB), to address it. Numerical results validate our algorithm's superiority over benchmark schemes, and simulations demonstrate that our algorithm achieves faster training speeds on the Fashion-MNIST and CIFAR-10 datasets within the same training duration.

\end{abstract}


\section{Introduction}
\IEEEPARstart{W}{ITH} the rapid development of industrial Internet of things (IoT) technology, a vast number of IoT devices are being installed in industrial facilities to enable data sensing and monitoring \cite{pokhrel2020efficient}. To enhance resource utilization efficiency in IoT networks and meet quality of service (QoS) standards, there is a need for dynamic and intelligent management of multi-dimensional resources like communication, computation, energy, and storage across the network. As an emerging technology, digital twin (DT) has been proposed to enhance the ability of managing various resources in the industrial IoT network scenarios from the perspective of the digital world \cite{duran2023digital}. DT creates a virtual representation in the digital realm by mapping physical entities with their features, state, and evolution \cite{lu2020communication, tao2018digital, li2023adaptive}. However, DT may face challenges in accurately modeling complexity and predicting unforeseen events in real time, which can be addressed through advanced machine learning techniques.

Utilizing machine learning models, DT facilitates intelligent decision-making in dynamic environments efficiently and effectively. For instance, in the industrial manufacturing sector, edge servers with DT can assess the connectivity performance of various industrial devices and proactively allocate communication resources to specific critical devices, thereby improving users' long-term QoS. Consequently, machine learning-empowered DT provides a pathway to achieve dependable network orchestration in wireless networks by modeling and forecasting the dynamic behavior of IoT networks. However, both DT and central machine learning rely on extensive data support dispersed across IoT devices. The existence of ``data silos", coupled with issues related to privacy and security, makes large-scale data collection impractical \cite{mu2023digital,lim2020federated}.

Federated Learning (FL) is considered as an efficient solution to address the aforementioned challenges, which is widely applied for building an intelligent model in a cost-efficient and privacy-preserving manner \cite{wei2020federated1, wei2022user, li2022blockchain, wei2022low, ma2023trusted}. 
FL significantly alleviates the burden of DT modeling by distributing model training across diverse IoT devices in parallel. In \cite{sun2020dynamic}, FL was integrated with DT to resolve the conflict between model training and privacy protection. 
The work \cite{mu2023digital} considers a communication-assisted sensing scenario in DT-empowered mobile networks utilizing FL. 
In \cite{lu2020low}, Lu \textit{et al.} proposed a new DT wireless network model based on FL and blockchain, enhancing the running efficiency of the model. The work in \cite{he2023client} introduced a new FL framework that integrates DT and proposed cluster-based client selection and DDPG-based resource allocation algorithm to improve performance and reduce delay in dynamic industrial IoT environments.

However, the above studies focused on integrating synchronous FL with DT, thereby overlooking the straggler effect in traditional FL, which can result in significant delays. Moreover, most of these studies did not adequately address the dynamic resource scheduling problem when network state information is unknown, often resulting in inefficient FL training. 
Although the work in \cite{he2023client} employed deep reinforcement learning (DRL) to tackle resource scheduling in wireless FL-based DT networks, the limitations of DRL are evident \cite{liu2020path}. 
On one hand, DRL's scalability is weak, as well-trained models often struggle to adapt to environmental changes, such as variations in the number of devices participating in FL training. On the other hand, training DRL models requires substantial computational resources, further complicating the implementation of asynchronous FL and DT.

Synchronous FL encounters several issues during implementation. First, all devices usually train their models simultaneously and upload the local models to the server, resulting in high instantaneous concurrent communication overhead. Second, for devices with slower training speeds or unstable network connections, synchronously uploading models for global aggregation poses a significant challenge. Lastly, due to slower training speeds or upload delays caused by network issues, other devices may have to wait for extended periods, leading to resource wastage and reduced training efficiency. To address these issues, a new training mechanism called asynchronous FL has been proposed. Hence, we aim to design efficient dynamic resource scheduling algorithms for lightweight DT-empowered edge networks wireless asynchronous FL. The main contributions of this paper are as follows:
\begin{enumerate}
  \item To tackle the issues of data silos and privacy security in industrial IoT, we develop a dynamic resource scheduling algorithm for the asynchronous FL to optimize the lightweight DT empowered IoT network. To enhance the global performance of asynchronous FL, we aim to minimize a multi-objective function that considers both energy consumption and latency by optimizing IoT device selection and transmission power control within an FL model performance constraint.
  \item Then, we utilize the Lyapunov method to decouple the formulated problem into a series of one-slot optimization problems. On the IoT device side, optimal close-form transmit power solutions are derived through a series of proofs and optimizations. On the edge server side, since partial state information is unknown, e.g., the transmitting power and computational frequency of IoT devices, the edge server leverages a multi-armed bandit (MAB) framework to model the IoT device selection problem and an efficient online algorithm, namely the client utility-based upper confidence bound (CU-UCB), to solve it. Furthermore, we theoretically derive the optimality gap, demonstrating sub-linear regret performance over communication rounds.
  \item Numerical results indicate that our proposed CU-UCB algorithm achieves a faster training speeds compared to other baseline algorithms on the Fashion-MNIST and CIFAR-10 datasets within the same training duration.
\end{enumerate}

The rest of the paper is organized as follows. In Section \ref{sec:related works}, the related works are discussed. In Section \ref{sec:system_model}, we describe the system model. The problem formulation and CU-UCB algorithm design is presented in Section \ref{Problem Formulation}. The performance of CU-UCB is shown in Section \ref{Simulations}. Finally, we conclude this paper in Section \ref{sec:conclusion}.

\section{Related Works}
\label{sec:related works}

In this section, we review the latest related work on FL for DT and resources-efficient asynchronous FL.

\subsection{Federated Learning for Digital Twin}
The DT paradigm stands out as one of the most promising technologies in mobile networks, enabling near-instant communication and highly reliable mobile services \cite{tao2018digital}. FL mitigates the risk of privacy exposure for clients by performing local training. Combining DT and FL in IoT networks is an emerging trend in academic research, leveraging FL for data protection and DT for efficient modeling \cite{mu2023digital}. For instance, \cite{jiang2021cooperative}  used FL technology to help resource-constrained mobile devices build DTs within wireless networks managed by different mobile network operators. In \cite{lu2020communication}, the authors proposed an FL framework for the DT-empowered IoT to enhance privacy protection, formulating the FL scheme as an optimization problem. Zhang et al. introduced FL into a DT-enabled Industrial IoT system to provide instant intelligence services for Industry 4.0 \cite{zhang2021energy}. In \cite{qian2023secrecy}, the authors investigated efficient data communication and computation in the construction of a FL-assisted marine DT network, addressing the challenges posed by eavesdropping attacks. Although the aforementioned studies explored the extensive applications of FL for DT, they overlooked the complexity of DT networks and failed to address the resource scheduling issues in FL for DT networks in dynamic environments.

\subsection{Resources-efficient Asynchronous Federated Learning}
Research on resource scheduling in asynchronous FL is still in its early stages, especially in industrial and DT scenarios, despite some studies exploring this area \cite{zhu2022online,lu2020communication1,xu2023energy}. For example, the authors in \cite{zhu2022online} proposed an asynchronous FL framework with adaptive client selection to minimize training latency, considering client availability and long-term fairness. However, this work focused solely on client scheduling, neglecting comprehensive issues such as communication, computational resources, and energy consumption in wireless asynchronous FL systems. Additionally, it did not optimize client-side resources. Moreover, the authors in \cite{zhu2022online} emphasized the fairness of client scheduling but overlooked the impact of resource scheduling on the performance of the global model. To enhance the communication efficiency of asynchronous FL in DT Edge Networks, \cite{lu2020communication1} proposed an asynchronous model update scheme and formulated the problem of reducing communication costs as an optimization problem. Nonetheless, this work solely addressed resource scheduling issues in static scenarios, ignoring the dynamic nature of asynchronous FL in DT Edge Networks. The work in \cite{xu2023energy} proposed a deep reinforcement learning-based energy-efficient algorithm to minimize energy consumption and accelerate model convergence, thereby enhancing the efficiency of asynchronous FL in mobile edge computing (MEC) networks by dynamically selecting mobile devices to participate in global aggregation. 

\section{System Model}
\label{sec:system_model}
\begin{figure}
\centering
\includegraphics[width=0.47\textwidth]{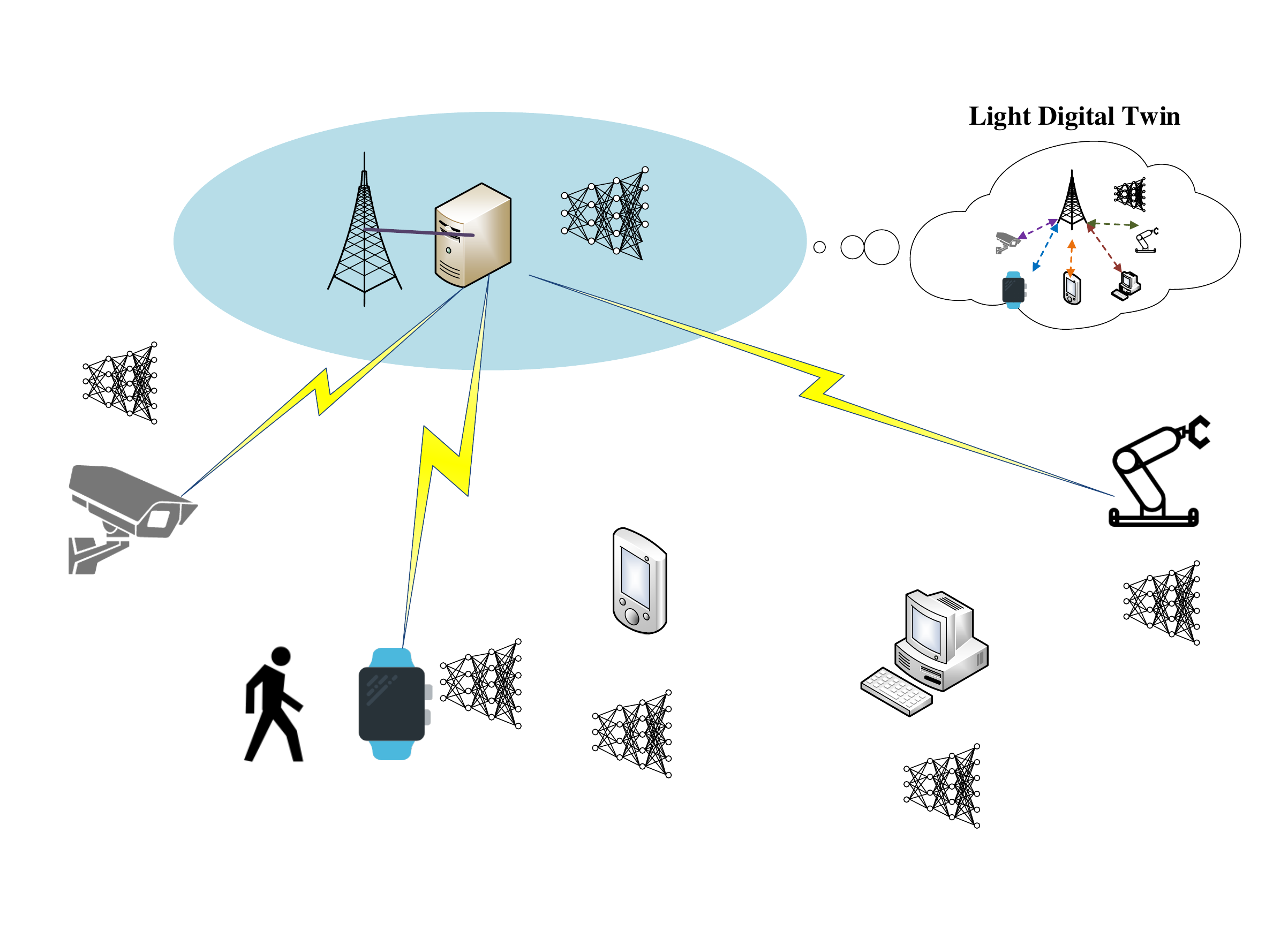}
\caption{The architecture of digital twin empowered IoT networks.}
\label{fig:system model figure}
\end{figure}
\subsection{Digital Twin for Internet of Things Networks}
A DT-empowered the wireless IoT network architecture as depicted in Fig. \ref{fig:system model figure}, comprising various types of IoT devices, an edge server, and a DT on the edge server. We employ the set $\mathcal{N}=\{1,2,...,N\}$ to represent the IoT devices within the network. The IoT network facilitates the interconnection of a wide array of IoT devices, sensors, and software, enabling the collection, analysis, and sharing of IoT data. Due to the extensive network scale, substantial data volumes, and device heterogeneity, IoT networks are highly challenging in terms of scheduling, decision-making, and analysis. DT can provide real-time status and dynamic topology mapping for the IoT devices.

DTs are virtual replicas of IoT devices, which are implemented in edge servers in our context. Nevertheless, constructing DT models often demands substantial real-time data, posing challenges for edge servers. A lightweight DT, designed specifically for capability-constrained devices, addresses this by providing a custom digital representation, which includes a state vector that maps key entity states and an intelligent analytics model with low computational complexity. 

The lightweight DT established on the edge server maps the key statuses (e.g., communication link conditions and resource availability) of the entire network in digital space and constructs a machine learning model to support intelligent network orchestration. The lightweight DT model can be denoted as
\begin{equation}
    DT=\{H(\mathcal{N}),S(\mathcal{N}), \mathcal{M}(\mathcal{N}) \}, \notag
\end{equation}
where $H(\mathcal{N})$ is the communication condition between IoT devices and the edge server, $S(\mathcal{N})$ is the key state of IoT devices that is related to intelligent network orchestration, and $\mathcal{M}(\mathcal{N})$ represents the machine learning model of intelligent network orchestration. It is noted that DT modelling relies on massive data, which cannot be provided by the edge server away from data. Besides, transferring local data from IoT devices to edge servers for constructing DT models can incur substantial communication overhead and pose privacy risks. The introduction of FL into lightweight DT provides an effective solution to the aforementioned issues. The lightweight DT based on FL can enhance both data privacy and modelling efficiency in the co-training of IoT devices, so that the edge server can make intelligent decisions to optimize IoT network orchestration.

\subsection{Asynchronous Federated Learning Based DT Modelling}
In this paper, we employ asynchronous FL for lightweight DT modelling. In contrast to synchronous FL, asynchronous FL provides enhanced flexibility and mitigates the impact of straggler issues. The edge server, acting as the aggregator of FL, distributes the DT model to IoT devices. The IoT devices sense data through their sensors, undergo local training with the acquired data, and subsequently transmit the local DT model to the edge server for aggregation.

Each IoT device possesses its own local dataset denoted as $\mathcal{D}_n$, consisting of $D_n=\left| \mathcal{D}_n \right|$ local samples $\{ \boldsymbol{x}_i\}_{i=1}^{D_n}$. For a given DT modelling parameter $\boldsymbol{w}$, the local loss function of a typical IoT device $n$ is $L_n(\boldsymbol{w})=\frac{1}{D_n} \sum_{\mathcal{D}_n} l(\boldsymbol{w})$, in which $l(\boldsymbol{w})$ means the loss function on a local sample. The global loss function is then expressed as $L(\boldsymbol{w})=\frac{1}{D}\sum_{n=1}^N D_n L_n(\boldsymbol{w})$, with $D$ denoting the total number of local data samples across all IoT devices. Similar to synchronous FL, asynchronous FL also aims to find the optimal weight $\boldsymbol{w}^*$, minimizing the global loss function $L(\boldsymbol{w})$. Specifically, the steps for a round of asynchronous FL are as follows,
\begin{itemize}
    \item \textbf{Global model publish:}  In each communication round, the edge server initially selects an IoT device $n$ from the available devices and subsequently distributes the global lightweight DT model to the selected device. 
    \item \textbf{Local model updates:} The selected IoT device $n$ solves the following local optimization problem, 
    \begin{equation}
        \label{SGD optimizer}
        \min_{\boldsymbol{w}} L_n(\boldsymbol{w})+\frac{m}{2}|| \boldsymbol{w}- \boldsymbol{w}(t-1)||^2,
    \end{equation}
    in which $m$ denotes the non-negative regularization constant, and $\boldsymbol{w}(t-1)$ means the global DT model parameters of the global round $t-1$.
     \item \textbf{Asynchronous global aggregation:} Once the edge server successfully receives the local DT model from the IoT device, it will initiate asynchronous global aggregation, 
     \begin{equation}
        \label{asynchronous global aggregation}
        \boldsymbol{w}(t)=(1-\rho(t))\boldsymbol{w}(t-1)+\rho(t)\boldsymbol{w}_n(t), \  \rho(t) \in (0,1),
    \end{equation}
    where $\rho(t)$ is the aggregation weight of the communication round $t$, and $w_n(t)$ means local model of IoT device $n$. According to \cite{xie2019asynchronous}, $\rho(t)=\rho \times s(t-\tau)$, in which $s(t-\tau)$ is the staleness function, evaluating the influence of IoT device $n$ on the global model. Typically, $t-\tau$ indicates the elapsed aggregation rounds at the IoT device since its last reception of an updated global model. In general, $s(t-\tau)$ should be $1$ when $t=\tau$, and monotonically decrease when $(t-\tau)$ increases. Hence, we employ the polynomial function that satisfy such two properties for the staleness function, i.e.,
    \begin{equation}
        \label{polynomial staleness function}
        s(t-\tau)=(t-\tau+1)^{-a}, \ a>0.
    \end{equation}
    The edge server and IoT devices conduct model updates asynchronously until the global DT modelling converges. The asynchronous FL-based DT model aggregation avoids long waiting times, which contributes to fast model learning and wait-free communication.
\end{itemize}

Here, we present two foundational assumptions for analyzing the convergence of asynchronous FL based lightweight DT modelling.
\begin{assumption}
\label{smooth function}
(Smoothness) The differentiable loss function $l(w)$ is $\beta$-smooth if for all $\boldsymbol{w}$, $\boldsymbol{w}'$, i.e., $l(\boldsymbol{w})-l(\boldsymbol{w}') \leq \langle\nabla l(\boldsymbol{w}'),\boldsymbol{w}-\boldsymbol{w}'\rangle+\frac{\beta}{2}||\boldsymbol{w}-\boldsymbol{w}'||^2$, in which $\beta>0$.
\end{assumption}

\begin{assumption}
\label{weak convexity}
(Weak convexity) The differentiable loss function $l(\boldsymbol{w})$ is termed $\mu$-weakly convex if the function $g(\boldsymbol{w})=l(\boldsymbol{w})+\frac{\mu}{2}||\boldsymbol{w}||^2$ exhibits convexity for all $\mu>0$.
\end{assumption}

Quoting to \cite{xie2019asynchronous}, given the two aforementioned assumptions, we can obtain the following property regarding the convergence guarantees of asynchronous FL. For brevity, we merely cite the convergence conclusions of asynchronous FL, while the comprehensive details and rigorous proofs can be found in \cite{xie2019asynchronous}.

\begin{theorem}
\label{convergence theorem}
We assume that the loss function $l$ satisfies {\it Assumption \ref{smooth function}} and {\it\ref{weak convexity}}, and each IoT device trains at most $\hat{D}_{\max}$ local samples before uploading models to the edge server. We also assume that the staleness is bounded by $\Delta$ for all IoT devices in any communication round, i.e., $\max_{n \in \mathcal{N}}\{t_n-\tau_n\} \leq \Delta$. Besides, we assume that $||\nabla l(\boldsymbol{w}_n;\boldsymbol{x}_i)||^2 \leq A_1$ and $||\nabla g_{\boldsymbol{w}'_n}(\boldsymbol{w}_n;\boldsymbol{x}_i)||^2 \leq A_2$, $\forall  \boldsymbol{x}_i \in \mathcal{D}_n$, $\forall n \in \mathcal{N}$. Last but not least, for any small $\varepsilon_0$, take $\zeta$ such that $\zeta > \mu$ and $(1+2\zeta+\varepsilon_0)A_2-\zeta^2 ||\boldsymbol{w}(\tau,d-1)-\boldsymbol{w}(\tau)||^2+\frac{\zeta}{2} ||\boldsymbol{w}(\tau,d-1)-\boldsymbol{w}(\tau)||^2 \leq 0$, in which $d \in [\hat{D}_{\min}, \hat{D}_{\max}]$ and $\boldsymbol{w}(\tau, d-1)$ is the model parameters derived from $\boldsymbol{w}(\tau)$ after $d-1$ rounds of local sample training. Thus, given a learning rate $\eta < \beta$, after $T$ communication rounds, the upper bound of convergence for the asynchronous FL based DT modelling can be expressed as follows:
\begin{equation}
\begin{split}
\label{upper bound of convergence}
        &\min_{t=1}^{T} \mathbb{E} ||\nabla L(\boldsymbol{w}(t))||^2 \leq \frac{\mathbb{E} [L(\boldsymbol{w}(1))-L(\boldsymbol{w}(T+1))]}{\rho \eta \varepsilon_0 T\hat{D}_{\min}}
        \\& +\frac{\eta T \hat{D}_{\max}^3}{\varepsilon_0 T\hat{D}_{\min}}\mathcal{O}(A_2)
        +\frac{\rho \Delta T \hat{D}_{\max}}{\varepsilon_0 T\hat{D}_{\min}} \mathcal{O}(\sqrt{A_1 A_2}) \\&+ \frac{\rho^2 \eta \Delta^2 T \hat{D}_{\max}^2}{\varepsilon_0 T\hat{D}_{\min}} \mathcal{O}(A_2) + \frac{\eta \Delta^2 T \hat{D}_{\max}^2}{\varepsilon_0 T\hat{D}_{\min}} \mathcal{O}(A_2).
\end{split}
\end{equation}
\end{theorem}


\begin{remark}
\label{remark1}
From \textbf{Theorem \ref{convergence theorem}}, we know that the upper bound of convergence for the asynchronous FL $\min_{t=1}^{T} \mathbb{E} ||\nabla L(\boldsymbol{w}(t))||^2$ is negatively correlated with $T\hat{D}_{\min}$, where $T\hat{D}_{\min}$ can be viewed as the total minimum number of training samples across all communication rounds. Inspired by this, in Section \ref{Problem Formulation}, we will consider the average minimum number of training samples as a long-term constraint to enhance the global performance of asynchronous FL.
\end{remark}

\subsection{Computation and Communication Models}
\label{Computation and Communication Models}
Typically, asynchronous FL comprises computational and communication processes. In the computation process, the chosen IoT device trains its local DT model using local datasets, resulting in updated local models. In the communication process, the trained IoT device uploads its local model parameters to the edge server for aggregation. Throughout these processes, IoT devices inevitably incur time and energy expenditures.

\textbf{Computation model:} We denote $f_n(t)$ as the computing resource contributed by IoT device $n$ for the local training process in communication round $t$, which is referred to as CPU frequency (in cycles/s). Note that IoT devices may also simultaneously handle multiple other computing tasks, thus, the CPU frequency of each IoT device varies in each communication round. The quantity of CPU cycles necessary for IoT device $n$ to process a single unit of dataset sample during local training is denoted by $c_n$, commonly referred to as the computational intensity of local training. The time consumption of IoT device $n$ for local training in communication round $t$ is expressed as
\begin{equation}
        \label{Time consumption of IoT device for local training}
        T^{\rm cmp}_n(t)=\frac{D_n c_n}{f_n(t)}.
\end{equation}

The energy consumption of IoT device $n$ for local training in communication round $t$ is denoted by
\begin{equation}
        \label{Energy consumption of IoT device for local training}
        E^{\rm cmp}_n(t)=\zeta T^{\rm cmp}_n(t) f_n(t)^3= \zeta D_n c_n f_n(t)^2,
\end{equation}
in which $\zeta$ is the effective capacitance parameter of computing chipset.

\textbf{Communication model:} The uplink transmission from the chosen IoT devices to the edge server is used to transmit their locally updated DT models, while the downlink is used to transmit the global DT models. Given the notably superior communication and computational capabilities of the edge server in contrast to IoT devices, we disregard the latency and energy consumption involved in the dissemination and aggregation of the global DT models in our paper. In communication round $t$, we express the transmit power of IoT device $n$ and the channel gain between IoT device $n$ and the edge server as $p_n(t)$ and $h_n(t)$, respectively. Inspired by \cite{xu2020task}, it is assumed that the channel gain remains constant within a communication round $t$ but varies across different communication rounds. Then, the uplink transmission rate used to upload the trained local model from the IoT device $n$ to the edge server in communication round $t$ can be expressed as
\begin{equation}
        \label{uplink transmission rate}
        r_n(t)=W \log_2\left( 1+ \frac{p_n(t) h_n(t)}{\delta_0^2}\right),\ \forall t.
\end{equation}
Here, we also assume that the wireless bandwidth in the system is evenly divided into  $M$ orthogonal subchannels, each with a bandwidth of $W$.

Each sub-channel can be allocated to at most one IoT device to avoid interference among them. Generally speaking, the available number $M$ of sub-channels is less than the total number $N$ of IoT devices in our system. The transmission time of IoT device $n$ in communication round $t$ is
\begin{equation}
        \label{the transmission time of IoT device n}
        T^{\rm com}_n(t)=\frac{z_n}{r_n(t)},
\end{equation}
in which $z_n$ represents the amount of data of the local model uploaded by IoT device $n$ (in bit). In general, owing to the requirement for edge servers to aggregate parameters uploaded by IoT devices, it is necessary for the local model parameters from each IoT device to have consistent specifications. In other words, the local model sizes of all IoT devices should be roughly the same. Hence, each IoT device selected for local model training is allocated the same bandwidth \cite{dynamic2021}. The energy consumption for uploading the local model of IoT device $n$ can be articulated as
\begin{equation}
        \label{the energy consumption for uploading the local model}
        E^{\rm com}_n(t)=p_n(t) T^{\rm com}_n(t).
\end{equation}
The complete process of updating a global DT model involves downloading the global DT model, conducting local model training, uploading the local model and aggregating the global DT model. Given the powerful communication and computational capabilities of edge servers, the latency associated with model downloading and aggregation is typically negligible. Hence, the total time spent by IoT device $n$ in a complete communication round can be expressed as
\begin{equation}
        \label{the total time spent by selected IoT device}
        T^{\rm tol}_n(t)=T^{\rm com}_n(t)+T^{\rm cmp}_n(t).
\end{equation}

The total energy consumption of IoT device $n$ in a complete communication round is calculated as
\begin{equation}
        \label{the total energy spent by selected IoT device}
        E^{\rm tol}_n(t)=E^{\rm com}_n(t)+E^{\rm cmp}_n(t).
\end{equation}

\section{Problem Formulation}
\label{Problem Formulation}
In the DT-empowered wireless IoT network, the edge server acts as an aggregator aiming to rapidly acquire aggregated well-trained models. To prevent IoT devices from becoming ``stragglers", we can minimize the time spent by IoT devices to complete global model updates in each communication round, thereby enhancing the training efficiency of asynchronous FL. Furthermore, we also aim to minimize the total energy consumption of all IoT devices for global model updates. Therefore, our main objective for this work is to optimize the transmit power and device selection to minimize latency and energy consumption of asynchronous FL in the IoT network. Let $\Omega_n(t)$ represent the cumulative energy consumption and latency of IoT device $n$ for local update in communication round $t$ and can be calculated as
\begin{equation}
        \label{the cumulative energy consumption and latency of IoT device}
        \Omega_n(t)=\lambda_t\frac{T^{\rm tol}_n(t)}{T_{\max}}+\lambda_e\frac{E^{\rm tol}_n(t)}{E_{\max}},
\end{equation}
where $\lambda_t+\lambda_e=1, \lambda_t \geq 0, \lambda _e \geq 0$ gives the trade-off between energy efficiency and latency, $T_{\max}$ and $E_{\max}$ represent the maximum latency and maximum energy consumption for DT model updates of any chosen IoT device in each communication round, respectively. Due to time-varying $h_n(t)$ and $f_n(t)$, $\forall n \in \mathcal{N}$, we need to design a dynamic scheduling mechanism in our IoT network to minimize the long-term cumulative energy consumption and latency of asynchronous FL. Thus, the optimization problem is expressed as follows:

\begin{align}
\label{the optimization problem 1}
  (\mathcal{P}0)   &\min_{p_n(t),x_n(t)} \frac{1}{T} \sum_{t=1}^{T} \sum_{n \in \mathcal{N}_A(t)} \mathbb{E} \left\{x_n(t) \Omega_n(t)\right\}\\
     \text{s.t.}\quad  &\text{C1:}\ \lim_{T \rightarrow \infty }  \frac{1}{T}\sum_{t=1}^{T} \mathbb{E} \{ x_n(t) D_n(t)\} \geq D_{\min},\notag\\
     &\text{C2:}\ 0 \leq p_n(t) \leq p_n^{\max}, \notag\\
     &\text{C3:}\ \sum_{n \in \mathcal{N}_A(t)} x_n(t)=1,  \notag\\
     &\text{C4:}\ x_n(t) \in \{0,1\},  \notag\\
     &\text{C5:}\ E^{\rm tol}_n(t) \leq E_{\max} , \notag\\
     &\text{C6:}\ T^{\rm tol}_n(t) \leq T_{\max} , \ \forall t \in \{1,...,T\}, \forall n \in \mathcal{N}_A(t), \notag
\end{align}
in which $\mathcal{N}_A(t)$ denotes the set of available IoT devices for asynchronous FL in communication round $t$. At the beginning of asynchronous FL, i.e., $t=0$, the edge server randomly selects $M$ IoT devices for local training subject to $M$ available sub-channels. Hence, our objective function (\ref{the optimization problem 1}) is the cumulative weighted sum of energy consumption and latency for IoT devices, commencing from the first communication round, i.e., $t=1$. $x_n(t)$ indicates whether IoT device $n$ is selected to execute local DT model update in communication round $t$. C1 ensures that each IoT device actively engages in model update training and specifies the minimum long-term average number of training sample $D_{\min}$ used by IoT devices for local training. This setting is necessary, as indicated by \textbf{Theorem \ref{convergence theorem}}, which demonstrates that the upper bound of convergence $\min_{t=0}^{T-1} \mathbb{E} ||\nabla L(w_t)||^2$ for asynchronous FL is inversely proportional to the minimum number of training samples $T \hat{D}_{\min}$. Moreover, C1 also ensures that all IoT devices can participate in local training, guaranteeing fairness in the training process. C2 ensures that the transmit power of each IoT device does not exceed its maximum power budget in each communication round. C3 and C4 imply that in each communication round, only one IoT device is selected to perform model updates. Finally, C5 and C6 indicate that in each communication round, the delay and energy consumption for model updates by IoT devices must not exceed the maximum upper bounds $T_{\max}$ and $E_{\max}$, respectively.

Problem $(\mathcal{P}0)$ is a long-term dynamic optimization problem, where the challenge arises from the expectation operations in constraint C1 and the objective function. Moreover, while the edge server can estimate the channel gains of IoT devices through channel estimations, the available CPU frequencies of these devices vary frequently, making it difficult to capture them. To address problem $(\mathcal{P}0)$, we employ a combination of the Lyapunov optimization method \cite{hu2023aoi,battiloro2022lyapunov} and MAB framework.

\subsection{Single-round Optimization Problem}
We can transform C1 into a queue stability constraint \cite{neely2010stochastic}. Thus, we introduce a virtual queue $Q_n(t)$ for IoT device $n$ with the following update equation,
\begin{equation}
\label{a virtual queue}
    Q_n(t+1)=\max\{Q_n(t)+D_{\min}-D_nx_n(t), 0\}.
\end{equation}
We use the virtual queue to transform the long-term constraint C1. The original problem can be transformed if the $Q_n(t)$ remains mean rate stable such that, 
\begin{equation}
    \lim_{t \rightarrow \infty} \frac{\mathbb{E} \{Q_n(t)\}}{t}=0.
\end{equation}

Then, we define the Lyapunov drift plus penalty function as
\begin{equation}
\label{Lyapunov drift plus penalty function}
    U(t)=\sum_{n \in \mathcal{N}_A(t)} \tilde{V} x_n(t) \Omega_n(t)-Q_n(t) D_n x_n(t),
\end{equation}
where $\tilde{V} \geq 0$ is a weight scalar used to balance the importance between the objective function and the virtual queue of C1. Thus, the original problem ($\mathcal{P}0$) can be transformed into the following single-round optimization problem, aimed at determining IoT device transmit power and device selection for communication round $t$,

\begin{align}
\label{the optimization problem 2}
  (\mathcal{P}1)   &\min_{p_n(t),x_n(t)}  U(t)\\
     \text{s.t.}\quad  &\text{C2:}\ 0 \leq p_n(t) \leq p_n^{\max}, \notag\\
     &\text{C3:}\ \sum_{n \in \mathcal{N}_A(t)} x_n(t)=1,  \notag\\
     &\text{C4:}\ x_n(t) \in \{0,1\},  \notag\\
     &\text{C5:}\ E^{\rm tol}_n(t) \leq E_{\max} , \notag\\
     &\text{C6:}\ T^{\rm tol}_n(t) \leq T_{\max} , \ \forall t \in \{1,...,T\}, \forall n \in \mathcal{N}_A(t). \notag
\end{align}
It is emphasized that by solving the single-round optimization problem ($\mathcal{P}1$) in communication round $t$ and updating $Q_n(t)$ according to (\ref{a virtual queue}), we can approach the optimal value of problem ($\mathcal{P}0$) while satisfying the constraint C1.

\subsection{Solution for Single-round Optimization Problem ($\mathcal{P}1$)}
\label{Solution for Single-round Optimization Problem}
According to the primal decomposition theory, the single-round optimization problem ($\mathcal{P}1$) can be decomposed into the following two subproblems:
\begin{align}
\label{the optimization problem 4}
  (\mathcal{P}11)   &\min_{p_n(t)}   \tilde{V} \left[\theta_t T^{tol}_n(t)+\theta_eE^{tol}_n(t) \right] \\
     \text{s.t.}\quad  &\text{C2, C5 and C6}, \notag
\end{align}
and
\begin{align}
\label{the optimization problem 3}
  (\mathcal{P}12)   &\min_{x_n(t)}  \sum_{n \in \mathcal{N}_A(t)} \tilde{V} x_n(t) \Omega_n^*(t)-Q_n(t) D_n x_n(t) \\
     \text{s.t.}\quad  &\text{C3 and C4}, \notag
\end{align}
in which the variable $\tilde{V} \Omega_n^*(t)$ in the IoT device selection subproblem ($\mathcal{P}12$) is the optimal value of the transmit power allocation subproblem ($\mathcal{P}11$). For convenience, we use $\theta_t$ and $\theta_e$ to represent $\frac{\lambda_t}{T_{\max}}$ and $\frac{\lambda_e}{E_{\max}}$, respectively. Since $(\mathcal{P}12)$ nests a problem regarding the variable $p_n(t)$, we first need to solve the subproblem ($\mathcal{P}11'$) to obtain the optimal value $\Omega_n^*(t)$ at IoT device side. By equivalently transforming constraints C5 and C6, we can obtain the following problem:
\begin{align}
\label{the optimization problem 5}
  (\mathcal{P}11')   &\min_{p_n(t)}   \tilde{V} \left[\theta_t T^{tol}_n(t)+\theta_eE^{tol}_n(t) \right] \\
     \text{s.t.}\quad  &\text{C2:}\ 0 \leq p_n(t) \leq p_n^{\max} \notag\\
     &\text{C7:}\ p_n(t) \leq \notag\\&-\frac{\text{LambertW} \left(-\frac{a_n(t) \delta_0^2}{h_n(t)} (\ln2) 2^{-\frac{a_n(t) \delta_0^2}{h_n(t)}}\right)}{a_n(t) \ln2}-\frac{\delta_0^2}{h_n(t)} , \notag\\
    &\text{C8:}\ p_n(t) \geq \frac{\delta_0^2}{h_n(t)} \left(2^{\frac{z_n}{W (T_{\max}-\frac{D_n c_n}{f_n(t)})}} -1 \right) ,  \notag\\ &\qquad \forall t \in \{1,...,T\}, \forall n \in \mathcal{N}_A(t),  \notag
\end{align}
in which $a_n(t)=\frac{z_n}{W(E_{\max}-\zeta D_n c_n f_n(t)^2)}$, and the $\text{LambertW}(\cdot)$ function is the inverse function of $f(y)=y e^y$ and cannot be expressed explicitly.

Proof: See Appendix \ref{Proof of P4}. \hfill $\blacksquare$

For the sake of convenience, we use $\tilde{p}_n^{\max}(t)$ and $\tilde{p}_n^{\min}(t)$ to represent $-\frac{\text{LambertW} \left(-\frac{a_n(t) \delta_0^2}{h_n(t)} (\ln2) 2^{-\frac{a_n(t) \delta_0^2}{h_n(t)}}\right)}{a_n(t) \ln2}-\frac{\delta_0^2}{h_n(t)}$ and $\frac{\delta_0^2}{h_n(t)} \left(2^{\frac{z_n}{W (T_{\max}-\frac{D_n c_n}{f_n(t)})}} -1 \right)$, respectively.

The complexity of problem ($\mathcal{P}11'$) stems from the term $p_n(t) T_n^{\text{com}}(t)$ in the objective function. To simplify the problem ($\mathcal{P}11'$), we introduce a variable $\Upsilon_n(t)$, which can be represented as follows,
\begin{equation}
\label{variable function}
   \Upsilon_n(t)=\frac{1}{W \log_2\left(1+\frac{p_n(t) h_n(t)}{\delta_0^2}\right)}.
\end{equation}
Thus, problem ($\mathcal{P}$11') can be rewritten as follows,
\begin{align}
\label{the optimization problem 6}
 & (\mathcal{P}11'')\  \min_{\Upsilon_n(t)}   \theta_t |\boldsymbol{w}_n| \Upsilon_n(t)+ \theta_t \frac{D_n c_n}{f_n(t)}+ \notag\\ &\frac{\theta_e \delta_0^2}{h_n(t)} \left( 2^{\frac{1}{W \Upsilon_n(t)}}-1  \right)|\boldsymbol{w}_n| \Upsilon_n(t) + \theta_e \zeta D_n c_n f_n(t)^2\\
     \text{s.t.} &\text{C9:}\ \Upsilon_n\left(\min\{p_n^{\max}, \tilde{p}_n^{\max}(t)\}\right) \leq \Upsilon_n(t) \leq \Upsilon_n(\tilde{p}_n^{\min}(t))\notag\\
      &\qquad \forall t \in \{1,...,T\}, \forall n \in \mathcal{N}.  \notag
\end{align}
To find the optimal solution for problem ($\mathcal{P}11''$), it is essential to investigate the convexity of its objective function. With a straightforward proof, we can derive the following theorem.
\begin{theorem}
\label{convex objective function}
    The objective function of problem ($\mathcal{P}11''$) is convex with respect to $\Upsilon_n(t) > 0$.
\end{theorem}

Proof: See Appendix \ref{Proof of convex objective function}. \hfill $\blacksquare$

Given that constraint C9 is a convex set, we can conclude that problem ($\mathcal{P}11''$) is a convex optimization problem. 
To solve problem ($\mathcal{P}11''$), we need to set the first derivative of equation (\ref{the optimization problem 6}) equal to $0$, i.e.,
\begin{equation}
\begin{split}
\label{first derivatives of objective function equal 0}
\frac{d(F_n(t))}{d(\Upsilon_n(t))}= \theta_t |\boldsymbol{w}_n|+\frac{\theta_e \delta_0^2}{h_n(t)} \left( 2^{\frac{1}{W \Upsilon_n(t)}}-1  \right)|\boldsymbol{w}_n|\\
-\frac{\theta_e \delta_0^2}{h_n(t)}  2^{\frac{1}{W \Upsilon_n(t)}} \frac{\ln 2 |\boldsymbol{w}_n|}{W \Upsilon_n(t)}=0.
\end{split}
\end{equation}
By solving equation (\ref{first derivatives of objective function equal 0}), we can obtain the global optimal solution for the objective function (\ref{the optimization problem 6}) in problem ($\mathcal{P}11''$). Based on this, we can further derive:
\begin{equation}
\begin{split}
\label{first derivatives of objective function equal 0_1}
2^{\frac{1}{W \Upsilon_n(t)}} \left(1- \frac{\ln 2}{W \Upsilon_n(t)} \right)=1-\frac{\theta_t h_n(t)}{\theta_e \delta_0^2}.
\end{split}
\end{equation}
Then, we need to determine the number of solutions for equation (\ref{first derivatives of objective function equal 0_1}). Through proof, we can establish the following theorem:
\begin{theorem}
\label{number of solutions for equation}
The solution to equation (\ref{first derivatives of objective function equal 0_1}) exists and is unique, defined as $\hat{\Upsilon}_n(t)$.
\end{theorem}

Proof: See Appendix \ref{Proof of number of solutions for equation}. \hfill $\blacksquare$

By defining the left-hand side of equation (\ref{first derivatives of objective function equal 0_1}) as $\Phi_n(\Upsilon_n(t)) = 2^{\frac{1}{W \Upsilon_n(t)}} \left(1- \frac{\ln 2}{W \Upsilon_n(t)} \right)$, we can conclude the following:
\begin{enumerate}
    \item When $\hat{\Upsilon}_n(t) \geq \Upsilon_n(\tilde{p}_n^{\min}(t))$, we have
    \begin{equation}
        \label{case1_1}
        \begin{split}
            \Phi_n(\Upsilon_n(\tilde{p}_n^{\min}(t))) \leq 1-\frac{\theta_t h_n(t)}{\theta_e \delta_0^2}.
        \end{split}
    \end{equation}
     
    In this case, the derivation of objective function (\ref{the optimization problem 6}) in problem $(\mathcal{P}11'')$ is less than $0$ in $\left[\Upsilon_n(\min\{p_n^{\max}, \tilde{p}_n^{\max}(t)\}), \Upsilon_n(\tilde{p}_n^{\min}(t))\right]$, which results in decreasing objective function. Under this circumstance, $\Upsilon_n(\tilde{p}_n^{\min}(t))$ is the optimal solution for problem $(\mathcal{P}11'')$.

    \item When $\hat{\Upsilon}_n(t) \leq \Upsilon_n\left(\min\{p_n^{\max}, \tilde{p}_n^{\max}(t)\}\right)$, we have
    \begin{equation}
        \label{case2_1}
        \begin{split}
            \Phi_n(\Upsilon_n\left(\min\{p_n^{\max}, \tilde{p}_n^{\max}(t)\}\right)) \geq 1-\frac{\theta_t h_n(t)}{\theta_e \delta_0^2}.
        \end{split}
    \end{equation}
     In this case, the derivative of the objective function (\ref{the optimization problem 6}) in problem $(\mathcal{P}11'')$ is larger than $0$ in $\left[\Upsilon_n(\min\{p_n^{\max}, \tilde{p}_n^{\max}(t)\}), \Upsilon_n(\tilde{p}_n^{\min}(t))\right]$, which results in increasing objective function. Under this circumstance, $\Upsilon_n\left(\min\{p_n^{\max}, \tilde{p}_n^{\max}(t)\}\right)$ is the optimal solution for problem $(\mathcal{P}11'')$.

    \item When $\Upsilon_n\left(\min\{p_n^{\max}, \tilde{p}_n^{\max}(t)\}\right) <\hat{\Upsilon}_n(t) < \Upsilon_n(\tilde{p}_n^{\min}(t))$, we have 
    \begin{equation}
        \label{case3_1}
        \begin{split}
            &\Phi_n(\Upsilon_n\left(\min\{p_n^{\max}, \tilde{p}_n^{\max}(t)\}\right)) \leq 1-\frac{\theta_t h_n(t)}{\theta_e \delta_0^2},\\
            &\Phi_n(\Upsilon_n(\tilde{p}_n^{\min}(t))) \geq   1-\frac{\theta_t h_n(t)}{\theta_e \delta_0^2}.
        \end{split}
    \end{equation}
    In this case, $\hat{\Upsilon}_n(t)$ is the optimal solution for problem $(\mathcal{P}11'')$.
\end{enumerate}

Hence, by considering (\ref{case1_1}), (\ref{case2_1}), and (\ref{case3_1}), we can derive the optimal solution for problem $(\mathcal{P}11'')$ as follows,
 \begin{equation}
        \label{all cases of optimal solution}
        \begin{split}
           \left\{
\begin{array}{lr}
\Upsilon^*_n(t)=\Upsilon_n(\tilde{p}_n^{\min}(t)), \\ \qquad \qquad\qquad \text{if}\quad\Phi_n(\Upsilon_n(\tilde{p}_n^{\min}(t))) \leq 1-\frac{\theta_t h_n(t)}{\theta_e \delta_0^2}\\
\Upsilon^*_n(t)=\Upsilon_n\left(\min\{p_n^{\max}, \tilde{p}_n^{\max}(t)\}\right), \\
 \text{if}\quad\Phi_n(\Upsilon_n\left(\min\{p_n^{\max}, \tilde{p}_n^{\max}(t)\}\right)) \geq 1-\frac{\theta_t h_n(t)}{\theta_e \delta_0^2}\\
\Phi_n(\Upsilon^*_n(t))=1- \frac{\theta_t h_n(t)}{\theta_e \delta_0^2}, \qquad \qquad \qquad \text{otherwise}.
\end{array}
\right.
        \end{split}
 \end{equation}
Note that, intuitively, it is difficult to directly obtain a solution for $\Phi_n(\Upsilon^*_n(t))=1- \frac{\theta_t h_n(t)}{\theta_e \delta_0^2}$ in (\ref{all cases of optimal solution}). By performing a series of elementary transformations on $\Phi_n(\Upsilon^*_n(t))=1- \frac{\theta_t h_n(t)}{\theta_e \delta_0^2}$ in (\ref{all cases of optimal solution}), and utilizing the LambertW function, we can derive a closed-form expression for $\Upsilon^*_n(t)$ from $\Phi_n(\Upsilon^*_n(t))=1- \frac{\theta_t h_n(t)}{\theta_e \delta_0^2}$ as follows:
\begin{equation}
    \label{closed-form expression for optimal solution}
    \begin{split}
    \Upsilon^*_n(t)=&\frac{\ln2}{W \left(1+ \text{LambertW} \left( (\frac{\theta_t h_n(t)}{\theta_e \delta_0^2} -1) 2 ^{-\frac{1}{\ln 2}}\right) \right)}\\
   & \text{if} \quad \Phi_n(\Upsilon_n\left(\min\{p_n^{\max}, \tilde{p}_n^{\max}(t)\}\right)) \leq \\&\qquad 1-\frac{\theta_t h_n(t)}{\theta_e \delta_0^2} \leq \Phi_n(\Upsilon_n(\tilde{p}_n^{\min}(t))).
    \end{split}
\end{equation}
By substituting $\Upsilon^*_n(t)$ into (\ref{variable function}), we can deduce the optimal transmit power $p_n^*(t)$.

\subsection{Decisions for IoT Devices Selection}
\label{Decisions for IoT Devices Selection}
We will address the IoT device selection problem in Problem ($\mathcal{P}12$) in this subsection. Through channel estimation, the channel gain between the IoT devices and the edge server can be obtained by the edge server. However, the CPU frequency of IoT devices is typically variable and unknown, and their transmission power has not yet been determined, presenting a significant challenge for the edge server in making accurate decisions based on this information. In this scenario, a method named MAB can be employed to address the IoT device selection problem in $(\mathcal{P}12)$.

Problem $(\mathcal{P}12)$, with unknown $f_n(t)$ and $p_n(t)$ for all $n \in \mathcal{N}$ can be viewed as a MAB problem. To be specific, the $N$ candidate IoT devices can be considered as $N$ arms. Since the computing capability and transmit power of each IoT device cannot be predicted by the edge server in advance, the cost incurred by each arm play is unknown beforehand. To address $(\mathcal{P}12)$, we must select the appropriate arm in each communication round to minimize the time-average expected cost. However, striking a balance between utilizing past experiences and exploring new strategies for the future has always been a significant challenge in MAB problems.
The Upper Confidence Bound (UCB) algorithm is a classical MAB algorithm employed to strike a balance between exploration and exploitation for resource allocation \cite{xia2020multi,xu2021online}. Following the explore-exploit strategy, the algorithm selects an arm (or action) at each communication round to minimize the cost. Thus, we design an efficient online learning based IoT devices selection algorithm by extending the classical UCB policy in our system. 

We denote $\Gamma_n(t)$ as the count of times that IoT device $n$ has been selected at the end of communication round $t$. In our system, we initialize $\Gamma_n(0)=0$ for any IoT device $n$, and we have $\Gamma_n(t)=\sum_{t'=1}^{t} x_n(t')$. For any $t>1$, $\Gamma_n(t)$ can be updated according to the following equation,
\begin{equation}
        \label{updated equation}
        \begin{split}
        \Gamma_n(t+1)=
           \left\{
\begin{array}{lr}
\Gamma_n(t)+1, \quad  &\text{if}\  x_n(t+1)=1,\\ 
\Gamma_n(t), \quad&\text{otherwise}.
\end{array}
\right.
        \end{split}
\end{equation}
Then, we define the sample mean of $\Omega_n(t)$ as follows:
\begin{equation}
    \label{samp means}
    \overline{s}_n(t)=\frac{\sum_{t'=1}^{t} x_n(t') \Omega_n(t')}{\Gamma_n(t)},
\end{equation}
in which $\Omega_n(t')$ is the cumulative energy consumption and latency of IoT device $n$ for local training at communication round $t'$, which can be viewed as the cost caused by the IoT devices $n$ in communication round $t'$. Besides, we set $\overline{s}_n(t)=0$ for the special case when $\Gamma_n(t)=0$. The UCB estimate of $\overline{s}_n(t)$ can be expressed as follows,
\begin{equation}
    \label{UCB estimate1}
    \tilde{s}_n(t)=\max \left\{ \overline{s}_n(t-1) - \sqrt{\frac{3 \ln{t}}{2 \Gamma_n(t-1)}}, 0\right\},
\end{equation}
in which a lower $\overline{s}_n(t-1)$ indicates that the IoT device exhibited better performance previously, serving as past experience to guide future decision-making. Meanwhile, $\sqrt{\frac{3 \ln{t}}{2 \Gamma_n(t-1)}}$ can be seen as an exploration of IoT devices that could potentially enhance the experience of asynchronous FL-based DT modelling in the future. Particularly, we set $\tilde{s}_n(t)=0$ in the case of $\Gamma_n(t-1)=0$. Then, we combine the Lyapunov optimization and UCB method to make a decision on IoT device selection in communication round $t$, selecting the arm that minimizes the weighted sum of $\tilde{s}_n(t)$ and $-Q_n(t) D_n$, i.e.,
\begin{equation}
    \label{select the minimal arm}
    n^{*}(t)=\arg\min_{n \in \mathcal{N}_A(t)} \left( \tilde{V} \tilde{s}_n(t)- Q_n(t) D_n \right).
\end{equation}
The detailed online learning algorithm for IoT device scheduling is referred to as CU-UCB and summarized in Algorithm \ref{Online Learning Algorithm for IoT Device Scheduling}. Once an IoT device is selected by the edge server, it optimizes its transmit power in each communication round according to (\ref{all cases of optimal solution}), (\ref{closed-form expression for optimal solution}), and (\ref{variable function}).

\begin{algorithm}
\label{Online Learning Algorithm for IoT Device Scheduling}
\caption{CU-UCB: Online Learning client utility based upper confidence bound Algorithm for IoT Device Scheduling}
\begin{algorithmic}[1]
\FOR{$t=\{1,...,T\}$}
\FOR{$n \in \mathcal{N}_A(t)$}
\IF{$\Gamma_n(t-1)>0$}
\STATE Obtain $\tilde{s}_n(t)$ from (\ref{UCB estimate1});
\ELSE
\STATE Set $\tilde{s}_n(t)=0$;
\ENDIF
\ENDFOR
\STATE Select the optimal IoT device $n^*(t)$ at the current communication round $t$ according to (\ref{select the minimal arm});
\STATE The selected IoT device $n^*(t)$ will determine its optimal transmit power $p_{n^*}(t)$ according to (\ref{all cases of optimal solution}), (\ref{closed-form expression for optimal solution}) and (\ref{variable function});
\STATE Update $\Gamma_{n^*}(t)$ according to (\ref{updated equation});
\STATE Update $\overline{s}_{n^*}(t)$ according to (\ref{samp means});
\STATE Update the queue $Q_n(t+1), \forall n \in \mathcal{N}$ for all IoT devices following (\ref{a virtual queue});
\ENDFOR
\end{algorithmic}
\end{algorithm}

\subsection{Algorithm Analysis}
It is observed that our CU-UCB algorithm has the same structure as the common UCB, and its computational complexity is $\mathcal{O}\left(\sum_{t=1}^T N_{A}(t)\log_2N_{A}(t)\right)$, where $N_{A}(t)=|\mathcal{N}_A(t)|$ represents the number of available IoT devices in each communication round. It should be noted that $\mathcal{O}\left(\sum_{t=1}^T N_{A}(t)\log_2N_{A}(t)\right)$ is less than $\mathcal{O}\left(T N\log_2 N\right)$.

Besides, we need to prove an upper bound on the time-average regret $R_{\pi}(T)$ under the policy $\pi$, which 
means the difference between the expected time-average cost obtained under policy $\pi$ that chooses an arm $n(t)$ in communication round $t$ and the optimal cost $s^{*}$. It is defined as follows:
\begin{equation}
    \label{upper bound on the time-average regret}
    R_{\pi}(T)=\frac{1}{T} \sum_{t=1}^T  \mathbb{E} \left\{\sum_{n \in \mathcal{N}_A(t)} x_n(t) \Omega_n(t) \right\}-s^{*}.
\end{equation}
Thus, we have the following theorem.
\begin{theorem}
\label{theorem of upper bound on the time-average LUCB regret}
    Under our CU-UCB algorithm, the time-average regret defined in (\ref{upper bound on the time-average regret}) has the following upper bound:
    \begin{equation}
    \label{upper bound on the time-average LUCB regret}
    R_{\text{CU-UCB}}(T) \leq \frac{N D_{\min}^2}{2\tilde{V}}+\frac{2 \sqrt{6NT\ln{T}}+4N}{T}.
\end{equation}
\end{theorem}
Proof: See Appendix \ref{proof of upper bound}. \hfill $\blacksquare$

\section{Simulations}
\label{Simulations}
In this section, we present simulation results to demonstrate the effectiveness and advantages of the proposed CU-UCB algorithm. In this simulation, we consider a circular area with a radius of $0.5$ kilometers, where the edge server is situated at the center and $N=30$ IoT devices are uniformly distributed throughout the area. We assume that there are $15$ available sub-channels, and each sub-channel has a bandwidth of 1 MHz. The channel fading model between the IoT device and the edge server encompasses both large-scale and small-scale fading. The large-scale fading model is given by $L(dB) = 128.1 + 37.6 \log_{10}(10^{-3} d)$, where $d$ represents the distance between the IoT device and the edge server in meters. Small-scale fading adheres to the normalized Rayleigh distribution, while the noise power density is set at $-154$ dBm for a $1$ MHz bandwidth.

In each communication round, the available CPU frequency at each IoT device $n$ follows a normal distribution with a mean of $\overline{f}_n$ and a standard deviation of $0.2*10^{9}$, in which $\overline{f}_n \in [10^9, 3*10^9]$. The maximum transmit power of the IoT device is set as $p_{\max}=1$ W. Besides, the maximum energy consumption $E_{\max}$ and the maximum delay are set as $1.2$ J and $1$ s, respectively. We initialize the $Q_n(1)=0$ in Algorithm \ref{Online Learning Algorithm for IoT Device Scheduling} for all IoT devices. The number of data samples for each IoT device is randomly selected from the range [70, 100]. Finally, the model size of lightweight DT modelling is set as $1$ MB.

To demonstrate the effectiveness of the CU-UCB algorithm, we utilize the ResNet-18 model to conduct image classification on 10 classes of images from CIFAR-10 and Fashion-MNIST datasets. As classic datasets in machine learning, Fashion-MNIST comprises 60,000 training samples and 10,000 test samples, while CIFAR-10 is a color image dataset closely representing everyday objects, consisting of 50,000 training images and 10,000 test images. We consider three baseline algorithms, i.e., As-Q-only, As-fairness, and Sy-fairness. As-Q-only is an algorithm crafted for dynamic environment, concentrating solely on ensuring queue stability and optimizing the transmission power of IoT devices\footnote{Through multiple simulation experiments, we found that the performance of the As-Q-only method on real-world datasets closely approaches that of the CU-UCB algorithm. This is because the As-Q-only algorithm primarily focuses on ensuring the long-term constraint C1 of asynchronous FL performance. However, the superiority of the CU-UCB algorithm over the As-Q-only method is more evident in terms of energy efficiency and latency, as shown in Fig. 3. Therefore, in subsequent simulations, we will only present the performance differences between the CU-UCB algorithm and the As-Q-only method concerning energy consumption and latency.}. As-fairness is an algorithm crafted for asynchronous FL settings, prioritizing fairness in the participation of all IoT devices in training, without optimizing transmission power. Sy-fairness is an algorithm designed for synchronous FL settings, focusing solely on ensuring fairness in the participation of all IoT devices in training. Furthermore, in the FL simulation of the real dataset, we have introduced a random selection algorithm that randomly chooses devices to participate in FL training.

\begin{figure}
\centering
\includegraphics[width=0.47\textwidth]{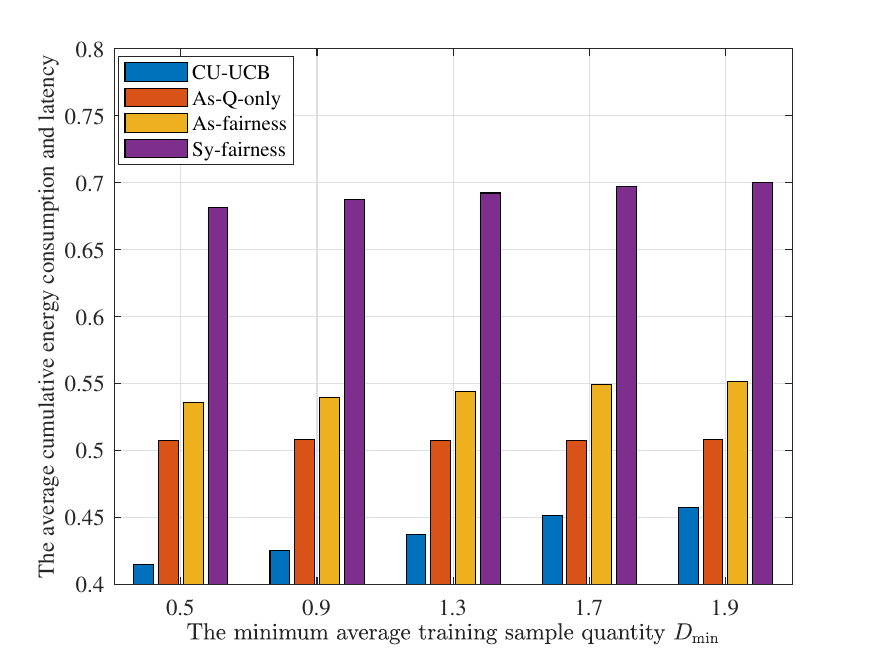}
\caption{The average cumulative energy consumption and latency versus the minimum average training sample quality $D_{\min}$ of the proposed CU-UCB and other baselines.}
\label{fig:D_minvsU}
\end{figure}

Fig.~\ref{fig:D_minvsU} depicts the average cumulative energy consumption and latency, i.e., the objective function of problem $(\mathcal{P}0)$ versus the minimum average training sample quality $D_{\min}$ of the proposed CU-UCB and other baselines. Compared to other baselines, our proposed CU-UCB algorithm demonstrates significant superiority across various values of $D_{\min}$. In contrast, the As-fairness algorithm exhibits the poorest performance, primarily due to the lower time efficiency of synchronous FL compared to asynchronous FL. Additionally, as $D_{\min}$ increases, the values of average cumulative energy consumption and latency also increase. This is because as the constraints strengthen, the feasible domain of problem $(\mathcal{P}0)$ continuously shrinks, leading to an increase in the minimum value of the objective function of $(\mathcal{P}0)$.

\begin{figure}
\centering
\includegraphics[width=0.47\textwidth]{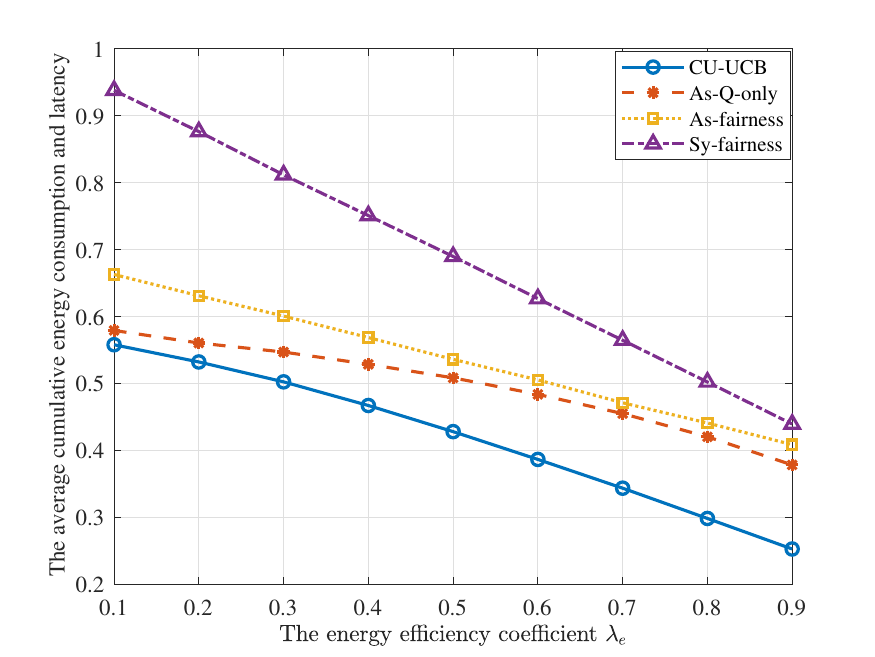}
\caption{The average cumulative energy consumption and latency versus the energy efficient coefficient $\lambda_{e}$ under the proposed CU-UCB and other baselines.}
\label{fig:lambda_U}
\end{figure}

Fig.~\ref{fig:lambda_U} shows the impact of the energy efficient coefficient $\lambda_{e}$ on the average cumulative energy consumption and latency for 4 different scheduling algorithms. As the energy efficiency coefficient $\lambda_{e}$ increases, the optimal value of average cumulative energy consumption and latency shows a decreasing trend. It is evident that our proposed CU-UCB algorithm consistently outperforms other baselines, both in terms of energy efficiency and time savings from Fig.~\ref{fig:lambda_U}. Furthermore, as $\lambda_{e}$ increases, the difference in average cumulative energy consumption and latency between the CU-UCB algorithm and the AS-Q-only algorithm continues to widen. This is because when $\lambda_{e}$ is small, the optimal average cumulative energy consumption and latency are more influenced by the power control strategy, whereas when $\lambda_{e}$ is large, the value is often more influenced by the scheduling strategy of IoT devices.

\begin{figure}
\centering
\includegraphics[width=0.47\textwidth]{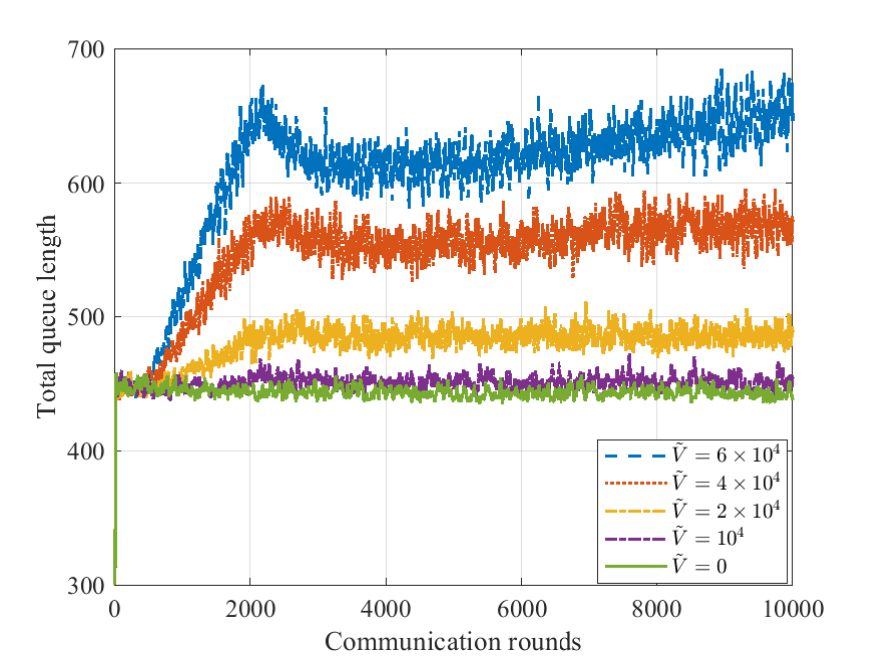}
\caption{Impact of $\tilde{V}$ on the total queue length of all IoT devices.}
\label{fig:virtual_queue}
\end{figure}

Fig.~\ref{fig:virtual_queue} depicts the influence of $\tilde{V}$ on the total queue length of CU-UCB with 10000 communication rounds of asynchronous FL and a total of 30 IoT devices. We observe that all the curves flatten out after a certain number of communication rounds. Intuitively, we notice that curves with larger  $\tilde{V}$ tend to converge more slowly, and the total queue length is also longer. The primary reason for this phenomenon is that larger values of $\tilde{V}$ tends to prioritize the optimization of the objective function in the early stages, with the stability of the long-term queue being considered afterwards.

\begin{figure}
\centering
\includegraphics[width=0.47\textwidth]{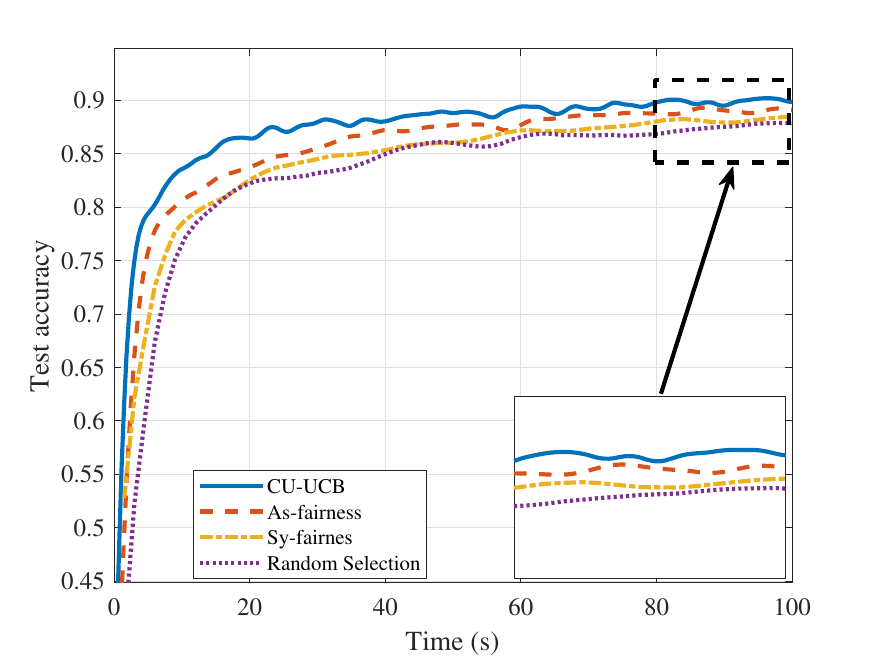}
\caption{Test accuracy versus training time for different schemes on the Fashion-MNIST dataset.}
\label{fig:fashion_mnist}
\end{figure}

\begin{figure}
\centering
\includegraphics[width=0.47\textwidth]{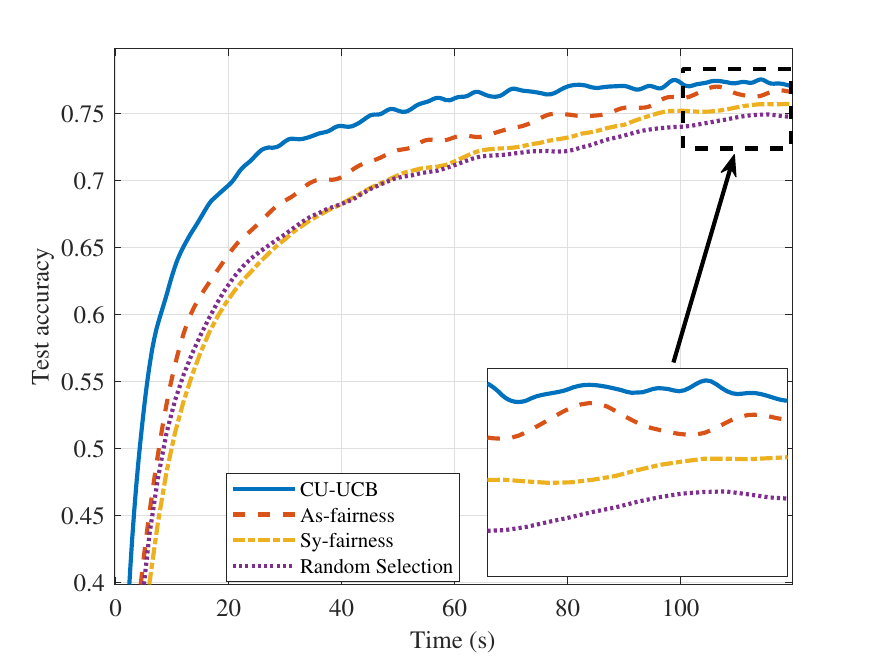}
\caption{Test accuracy versus training time for different schemes on the CIFAR-10 dataset.}
\label{fig:cifar_10_IID}
\end{figure}

Fig.~\ref{fig:fashion_mnist} and Fig.~\ref{fig:cifar_10_IID} depict the test accuracy versus the training time of the proposed CU-UCB and the baselines on the IID Fashion-MNIST dataset and the IID CIFAR-10 dataset. It can be observed that CU-UCB scheme achieves a better test accuracy performance than other algorithms. This is because, compared to other algorithms, the CU-UCB algorithm can enhance time utilization efficiency during training by efficiently scheduling resources. However, Sy-fairness algorithm must wait for the slowest IoT device to respond or for a timeout signal before aggregation, which often leads to wastage of wireless resources and computing capabilities of other IoT devices.

\begin{figure}
\centering
\includegraphics[width=0.47\textwidth]{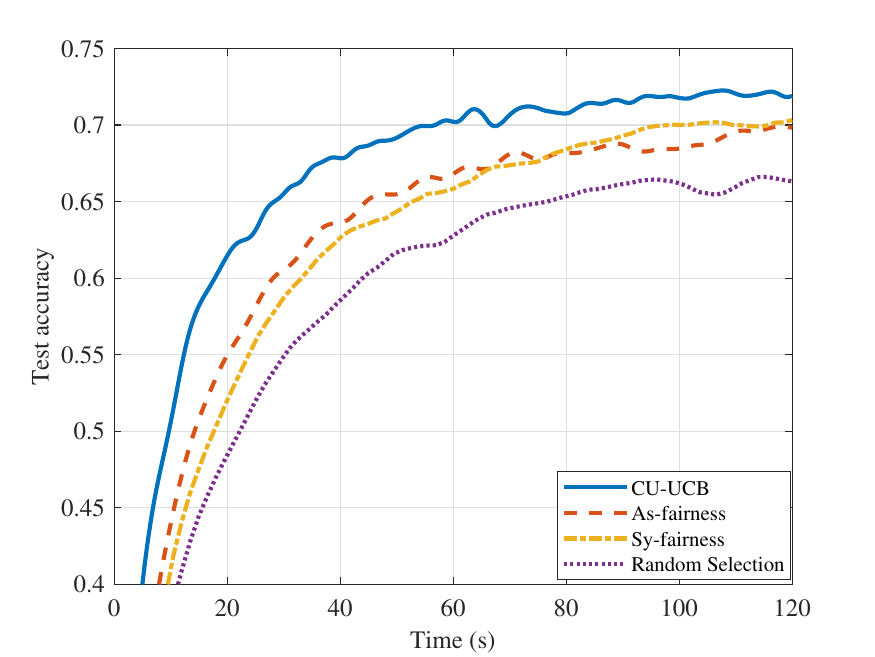}
\caption{Test accuracy versus training time for different schemes on the non-IID CIFAR-10 dataset.}
\label{fig:cifar10_noniid}
\end{figure}

In Fig.~\ref{fig:cifar10_noniid}, we explore the training performance in the non-IID setting. In our experiment, we use the same methodology as described in \cite{hsu2019measuring} for synthesizing non-identical user data. To be more precise, we uniformly extract $q_i \times 500$ elements from each class in our task. The vector $\boldsymbol{q}=(q_1,...,q_i)$ is generated from a Dirichlet distribution, denoted as $\boldsymbol{q} \sim Dir(\gamma_1 \boldsymbol{P})$. Here, $\boldsymbol{P}$ represents a 10-dimensional vector with each element equal to 1, and $\gamma_1$ is a concentration parameter that controls the degree of similarity between IoT device's data. If $\gamma_1 \to 0$, each device has only one randomly chosen class. Inversely, when $\gamma_1 \to \infty$, all devices have identical degrees to all classes (approximating iid data). It is evident that the test accuracy performance of the CU-UCB scheme under the non-IID setting still surpasses that of other baseline algorithms, although its performance is not as good as that under the IID setting as shown in Fig.~\ref{fig:cifar_10_IID}. This is because of the significant diversity among the data across IoT devices, making it more challenging for models to generalize to new data samples under non-IID data distributions.

\begin{figure}
\centering
\includegraphics[width=0.47\textwidth]{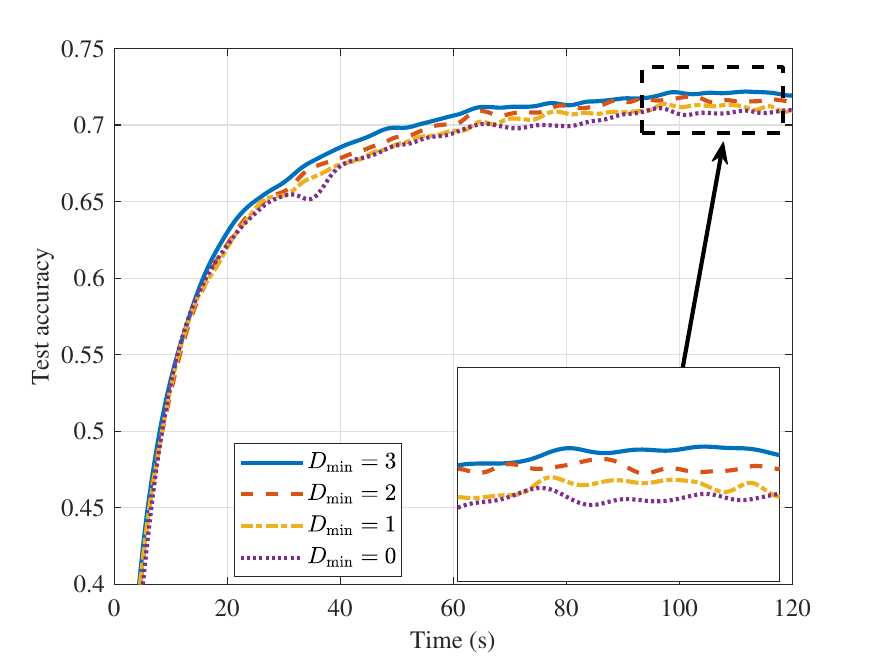}
\caption{Test accuracy versus training time for different minimum number of training samples $D_{\min}$ on the non-IID CIFAR-10 dataset.}
\label{fig:acc_D_min}
\end{figure}

Fig.~\ref{fig:acc_D_min} illustrates the impact of the minimum number of training samples, $D_{\min}$, on the test accuracy versus training time under the CU-UCB algorithm in the non-IID setting. It is evident that when $D_{\min}$ is larger, the test performance of asynchronous FL under the same settings is better, which aligns with the conclusion of \textbf{Theorem \ref{convergence theorem}} and \textbf{Remark \ref{remark1}}. In fact, a larger $D_{\min}$ indicates that each IoT device participates more evenly in asynchronous FL, which generally benefits the overall performance of the global model.

\section{Conclusion}
\label{sec:conclusion}
In this paper, we developed a dynamic resource scheduling algorithm for the asynchronous FL-based lightweight DT-empowered IoT network. Specifically, we aimed to minimize a multi-objective
function, capturing both the energy consumption and latency, by optimizing IoT device selection and transmit power control under an FL model performance constraint. Then, we utilized the Lyapunov method to decouple the formulated problem into a series of one-slot optimization problems. We optimized the transmit power for the chosen IoT device, achieving closed-form solutions using the LambertW function and convex optimization. Additionally, we proposed an online learning algorithm named CU-UCB for IoT device selection, which achieved sub-linear regret performance over communication rounds. Numerical results demonstrated the superiority of our algorithms compared to benchmark schemes. The simulation results also indicated that our CU-UCB algorithm achieved a faster training speed compared to the other baseline algorithms on the Fashion-MNIST dataset and the CIFAR-10 dataset under the same training duration.

%


\appendix
\subsection{Proof of ($\mathcal{P}4$)}
\label{Proof of P4}
According to (\ref{Time consumption of IoT device for local training}) and (\ref{the transmission time of IoT device n}), constraint C6 can be represented as
\begin{equation}
\label{C11_1}
\begin{split}
    \frac{z_n}{W \log_2\left( 1+\frac{p_n(t) h_n(t)}{\delta_0^2} \right)} + \frac{D_n c_n}{f_n(t)} \leq T_{\max}, \\ \forall t \in \{1,...,T\}, \forall n \in \mathcal{N}_A.
\end{split}  
\end{equation}
Through a series of basic arithmetic operations, (\ref{C11_1}) can be transformed into the following inequality,
\begin{equation}
\begin{split}
\label{C11_2}
    \frac{\delta_0^2}{h_n(t)} \left(2^{\frac{z_n}{W (T_{\max}-\frac{D_n c_n}{f_n(t)})}} -1 \right)\leq p_n(t), \\ \forall t \in \{1,...,T\}, \forall n \in \mathcal{N}_A.
\end{split}
\end{equation}
Correspondingly, based on (\ref{the energy consumption for uploading the local model}) and (\ref{the total energy spent by selected IoT device}), constraint C5 can be represented as follows,
\begin{equation}
\label{C10_1}
\begin{split}
    p_n(t)\frac{z_n}{W \log_2\left( 1+\frac{p_n(t) h_n(t)}{\delta_0^2} \right)} \leq E_{\max}- \zeta D_n c_n f_n(t)^2, \\ \forall t \in \{1,...,T\}, \forall n \in \mathcal{N}_A.
\end{split}
\end{equation}
For ease of analysis, we define $g_n(p_n(t))$ as follows,
\begin{equation}
\label{g(p_n)}
\begin{split}
    g_n(p_n(t))=p_n(t) \frac{z_n}{W \log_2 \left(1+ \frac{p_n(t)h_n(t)}{\delta^2_0} \right)}.
\end{split}
\end{equation}
By taking the first derivative of (\ref{g(p_n)}), we can obtain
\begin{equation}
\label{first derivative of g(p_n)}
\begin{split}
    &\frac{d \left(g_n(p_n(t))\right)}{d\left( p_n(t)\right)}
    =\frac{z_n \ln 2}{W \ln \left(1+ \frac{p_n(t)h_n(t)}{\delta^2_0} \right)}\\
    &\qquad\qquad -\frac{\ln 2 z_n p_n(t)}{W \ln^2 \left(1+ \frac{p_n(t)h_n(t)}{\delta^2_0} \right)} \cdot \frac{\frac{h_n(t)}{\delta^2_0}}{1+\frac{p_n(t)h_n(t)}{\delta_0^2}}\\
   &=\frac{\ln 2 z_n}{W \ln^2 \left(1+ \frac{p_n(t)h_n(t)}{\delta^2_0} \right)} \Bigg(\ln \left(1+\frac{p_n(t) h_n(t)}{\delta_0^2}\right)\\
    &\qquad \qquad-\frac{\frac{h_n(t)}{\delta_0^2} p_n(t)}{1+\frac{p_n(t) h_n(t)}{\delta^2_0}} \Bigg).
\end{split}
\end{equation}
We cannot directly determine the positive or negative characteristics from (\ref{first derivative of g(p_n)}), and consequently, it is difficult to assess the monotonicity of (\ref{g(p_n)}). To investigate the positivity or negativity of (\ref{first derivative of g(p_n)}), we define the following function $\tilde{g}(p_n(t))$,
\begin{equation}
\label{g(p_n)1}
\begin{split}
   \tilde{g}(p_n(t))=\ln \left(1+\frac{p_n(t) h_n(t)}{\delta_0^2}\right)
    -\frac{\frac{h_n(t)}{\delta_0^2} p_n(t)}{1+\frac{p_n(t) h_n(t)}{\delta^2_0}}.
\end{split}
\end{equation}
Thus, by taking the derivative of (\ref{g(p_n)1}), we can obtain
\begin{equation}
\label{first derivative of g(p_n)1}
\begin{split}
   \frac{d (\tilde{g}(p_n(t)))}{d(p_n(t))}=\frac{\left(\frac{h_n(t)}{\delta_0^2}\right)^2 p_n(t)}{\left[ 1+\frac{h_n(t)}{\delta_0^2}p_n(t)\right]^2} \geq 0.
\end{split}
\end{equation}
Due to $\tilde{g}(0)=0$ and $\frac{d (\tilde{g}(p_n(t)))}{d(p_n(t))} \geq 0$ for all $p_n(t) \geq 0$, $\frac{d \left(g_n(p_n(t))\right)}{d\left( p_n(t)\right)} \geq 0$ holds true for all $p_n(t) \geq 0$. Furthermore, we can infer that $g_n(p_n(t))$ is a monotonically increasing function with respect to $p_n(t)\geq 0$. Therefore, when $g_n(p_n(t))=E_{\max}-\zeta D_n c_n f_n(t)^2$, $p_n(t)$ achieves its maximum value. Through a series of elementary operations, the maximum transmit power $\tilde{p}_n^{\max}(t)$ of IoT device $n$ at communication round $t$ in C7 can be expressed as
\begin{equation}
\label{the maximum transmit power in C10}
\begin{split}
   &\tilde{p}_n^{\max}(t)=\\&-\frac{\text{LambertW} \left(-\frac{a_n(t) \delta_0^2}{h_n(t)} (\ln2) 2^{-\frac{a_n(t) \delta_0^2}{h_n(t)}}\right)}{a_n(t) \ln2}-\frac{\delta_0^2}{h_n(t)},
\end{split}
\end{equation}
in which $a_n(t)=\frac{z_n}{W(E_{\max}-\zeta D_n c_n f_n(t)^2)}$, and the $\text{LambertW}(\cdot)$ function is the inverse function of $f(y)=y e^y$, which cannot be expressed explicitly. \hfill $\blacksquare$

\subsection{Proof of Theorem \ref{convex objective function}}
\label{Proof of convex objective function}
First, we denote the objective function of ($\mathcal{P}11''$) as $F_n(t)=\theta_t z_n \Upsilon_n(t)+ \theta_t \frac{D_n c_n}{f_n(t)}+ \frac{\theta_e \delta_0^2}{h_n(t)} \left( 2^{\frac{1}{W \Upsilon_n(t)}}-1  \right)z_n \Upsilon_n(t) + \theta_e \zeta D_n c_n f_n(t)^2$. Thus, we respectively take the first and second derivatives of $F_n(t)$ as follows,
\begin{equation}
\begin{split}
\label{first derivatives of objective function}
\frac{d(F_n(t))}{d(\Upsilon_n(t))}= \theta_t z_n+\frac{\theta_e \delta_0^2}{h_n(t)} \left( 2^{\frac{1}{W \Upsilon_n(t)}}-1  \right)z_n\\
-\frac{\theta_e \delta_0^2}{h_n(t)}  2^{\frac{1}{W \Upsilon_n(t)}} \frac{\ln 2 z_n}{W \Upsilon_n(t)},
\end{split}
\end{equation}

\begin{equation}
\begin{split}
\label{second derivatives of objective function}
\frac{d^2(F_n(t))}{d^2(\Upsilon_n(t))}= \frac{\theta_e \delta_0^2}{h_n(t)} 2^{\frac{1}{W \Upsilon_n(t)}} \frac{(\ln^2 2) z_n}{W^2 \Upsilon^3_n(t)}.
\end{split}
\end{equation}
For all $\Upsilon_n(t))$, the second derivative (\ref{second derivatives of objective function}) is positive. Thus, the objective function of problem ($\mathcal{P}11''$) is convex with respect to $\Upsilon_n(t) > 0$. \hfill $\blacksquare$

\subsection{Proof of Theorem \ref{number of solutions for equation}}
\label{Proof of number of solutions for equation}
First, we define the left-hand side of equation (\ref{first derivatives of objective function equal 0_1}) as $\Phi_n(\Upsilon_n(t))=2^{\frac{1}{W \Upsilon_n(t)}} \left(1- \frac{\ln 2}{W \Upsilon_n(t)} \right)$. By computing the first derivative of $\Phi_n(\Upsilon_n(t))$, we obtain the following equation,
\begin{equation}
\begin{split}
\label{first derivatives of Phi}
\frac{d(\Phi_n(\Upsilon_n(t)))}{d(\Upsilon_n(t))}= 2^{\frac{1}{W \Upsilon_n(t)}}\frac{(\ln^2 2) }{W^2 \Upsilon^3_n(t)},
\end{split}
\end{equation}
where, when $\Upsilon_n(t)$ is greater than 0, $\frac{d(\Phi_n(\Upsilon_n(t)))}{d(\Upsilon_n(t))}$ consistently remains greater than 0. In other words, $\Phi_n(\Upsilon_n(t))$ monotonically increases within the interval $\Upsilon_n(t) \in (0, +\infty)$. Then, we need to investigate the values of $\Phi_n(\Upsilon_n(t))$ within the interval $\Upsilon_n(t) \in (0, +\infty)$. We respectively compute the limit of function $\Phi_n(\Upsilon_n(t))$ as $\Upsilon_n(t)$ approaches $0^+$ and as $\Upsilon_n(t)$ approaches $+\infty$, i.e.,
\begin{equation}
\begin{split}
\label{limit to 0}
&\lim_{\Upsilon_n(t) \rightarrow 0^+}\Phi_n(\Upsilon_n(t))\\&= \lim_{\Upsilon_n(t) \rightarrow 0^+} 2^{\frac{1}{W \Upsilon_n(t)}} \left(1- \frac{\ln 2}{W \Upsilon_n(t)} \right)=-\infty,
\end{split}
\end{equation}

\begin{equation}
\begin{split}
\label{limit to infty}
&\lim_{\Upsilon_n(t) \rightarrow +\infty}\Phi_n(\Upsilon_n(t))\\&= \lim_{\Upsilon_n(t) \rightarrow +\infty} 2^{\frac{1}{W \Upsilon_n(t)}} \left(1- \frac{\ln 2}{W \Upsilon_n(t)} \right)=1.
\end{split}
\end{equation}
Hence, for $\Upsilon_n(t) \in (0, +\infty)$, the function $\Phi_n(\Upsilon_n(t))$ is  is monotonically increasing, and $\Phi_n(\Upsilon_n(t)) \in (-\infty,1)$. In addition, due to $\frac{\theta_t h_n(t)}{\theta_e \delta_0^2} >0$, the right-hand side of equation (\ref{first derivatives of objective function equal 0_1}), i.e., $1-\frac{\theta_t h_n(t)}{\theta_e \delta_0^2} < 1$. Therefore, for any given $1-\frac{\theta_t h_n(t)}{\theta_e \delta_0^2}$, the function $\Phi_n(\Upsilon_n(t))$ has a unique $\hat{\Upsilon}_n(t)$ such that $\Phi_n(\hat{\Upsilon}_n(t))=1-\frac{\theta_t h_n(t)}{\theta_e \delta_0^2}$. Consequently, we can conclude that equation (\ref{first derivatives of objective function equal 0_1}) has a unique solution. \hfill $\blacksquare$

\subsection{Proof of Theorem \ref{theorem of upper bound on the time-average LUCB regret}}
\label{proof of upper bound}
We assume that $\pi^{*}$ is the best policy in our system. Let $\mathcal{N}^{*}(t)$ be the arms can be selected by policy $\pi^{*}$ in communication round $t$, and let $\boldsymbol{x^*}(t)=[x_1^*(t),...,x_N^*(t)]$ be the corresponding action vector. According to (\ref{upper bound on the time-average regret}), we can rewrite the regret of CU-UCB as
\begin{equation}
\label{rewrite the regret of LUCB}
R_{\rm LUCB}(T)=\frac{1}{T} \sum_{t=1}^T \mathbb{E} [\Delta R(t)],
\end{equation}
where
\begin{equation}
\Delta R(t) \triangleq \sum_{n\in \mathcal{N}(t)} x_n(t) \Omega_n(t)- \sum_{n\in \mathcal{N}^{*}(t)} x_n^*(t) \Omega_n(t).
\end{equation}

The Lyapunov function can be defined as $L(\boldsymbol{Q}(t)) \triangleq \frac{1}{2}\sum_{n=1}^N Q_n^2(t)$. According to the Lyapunov-drift analysis, the expected drift-plus-regret can be bounded as follows,
\begin{equation}
\label{bound of expected drift-plus-regret}
\begin{split}
&\mathbb{E}[L(\boldsymbol{Q}(t+1))-L(\boldsymbol{Q}(t))]+ \tilde{V} \mathbb{E}[\Delta R(t)] \leq  \frac{D_{\min}^2 N}{2}+\\ &\mathbb{E}\left[\sum_{n=1}^N (-Q_n(t)D_n + \tilde{V} \Omega_n(t))(x_n(t)-x_n^*(t)) \right]
\end{split}
\end{equation}
Summing (\ref{bound of expected drift-plus-regret}) over $t \in \{1,...,T\}$, applying the telescoping sum technique, and dividing both sides of the inequality by $\tilde{V} T$, we obtain the bound of regret for CU-UCB as
\begin{equation}
\label{upper bound2}
\begin{split}
&R_{\rm LUCB}(T) \leq \frac{D_{\min}^2 N}{2\tilde{V}}+ \\ &\frac{1}{T\tilde{V}} \sum_{t=1}^T \mathbb{E}\left[\sum_{n=1}^N (-Q_n(t)D_n+ \tilde{V} \Omega_n(t))(x_n(t)-x_n^*(t)) \right],
\end{split}
\end{equation}
in which the reason why the inequality holds in (\ref{upper bound2}) is due to the conditions $L(\boldsymbol{Q}(1))=0$ and $L(\boldsymbol{Q}(T))\geq 0$. Then, we consider another policy $\pi'$, and the rules for this policy are as follows,
\begin{equation}
\label{rule of new ploicy}
\begin{split}
n'(t) = \arg \min_{n\in \mathcal{N}'(t)} -Q_n(t)D_n+ \tilde{V} \Omega_n(t).
\end{split}
\end{equation}
Accordingly, we have $-Q_{n(t)}D_n+\tilde{V} \tilde{s}_{n(t)} \leq -Q_{n'(t)}D_{n'}+\tilde{V} \tilde{s}_{n'(t)}$ and 
\begin{equation}
\begin{split}
    &\sum_{t=1}^T \mathbb{E}\left[\sum_{n=1}^N (-Q_n(t)D_n+ \tilde{V} \Omega_n(t))(x_n(t)-x_n^*(t)) \right]\\
    &\leq  \sum_{t=1}^T \mathbb{E}\left[ (-Q_n(t)D_n+ \tilde{V} \Omega_n(t))\right.\\ 
    &\qquad \qquad \left.- (-Q_{n'}(t)D_{n'}+ \tilde{V} \Omega_{n'}(t))  \right]\\
    &\leq  \sum_{t=1}^T \mathbb{E}\left[ (-Q_n(t)D_n+ \tilde{V} \Omega_n(t))\right.\\ 
    &- (-Q_{n'}(t)D_{n'}+ \tilde{V} \Omega_{n'}(t)) + (-Q_{n'(t)}D_{n'}+\tilde{V} \tilde{s}_{n'}(t))\\
    &\qquad \qquad \left.- (-Q_{n(t)}D_n+\tilde{V} \tilde{s}_{n(t)}) \right]\\
    & = \tilde{V} \underbrace{\sum_{t=1}^T  \mathbb{E} \left[\Omega_n(t)- \tilde{s}_{n(t)} \right]}_{B1}+\tilde{V} \underbrace{\sum_{t=1}^T  \mathbb{E} \left[  \tilde{s}_{n'}(t)-\Omega_{n'}(t) \right]}_{B2}.
\end{split}
\end{equation}
According to the Chernoff-Hoeffding bound and the notion  of regret in typical MAB problems \cite{li2019combinatorial}, we can obtain the following upper bound,
\begin{equation}
\begin{split}
    B1 \leq 2 &\sqrt{6 N T \ln{T}} + \frac{5}{2}N,\\
    &B2 \leq \frac{3}{2}N.
\end{split}
\end{equation}
\hfill $\blacksquare$

%

\ifCLASSOPTIONcaptionsoff
\newpage
\fi


\bibliographystyle{IEEEtran}
\bibliography{ANN}





\end{document}